\documentclass[%
reprint,
superscriptaddress,
amsmath,amssymb,showkeys,
pra,
floatfix,
]{revtex4-2}

\usepackage[utf8]{inputenc}
\usepackage[linesnumbered,ruled,vlined]{algorithm2e}
\let\oldnl\nl
\newcommand{\nonl}{\renewcommand{\nl}{\let\nl\oldnl}}
\usepackage{braket}
\usepackage{float}
\usepackage{orcidlink}

\usepackage{graphicx}
\usepackage{dcolumn}
\usepackage{enumitem}
\usepackage{lipsum}  
\usepackage{bm}
\usepackage{hyperref}
\hypersetup{
    colorlinks=true,
    linkcolor=blue,
    citecolor=magenta
}
\usepackage{subcaption}
\begin{document}

\preprint{APS/123-QED}

\title{Surface Code Design for Asymmetric Error Channels}

\author{Utkarsh Azad\,\orcidlink{0000-0001-7020-0305}}
\email{utkarsh.azad@research.iiit.ac.in}
\affiliation{%
    Center for Computational Natural Sciences and Bioinformatics, International Institute of Information Technology, Hyderabad
}%
\affiliation{%
    Center for Quantum Science and Technology, International Institute of Information Technology, Hyderabad
}%
\affiliation{%
    Xanadu, Toronto, ON, M5G 2C8, Canada
}%
\author{Aleksandra Lipi\'{n}ska}
\affiliation{%
    Faculty of Mathematics and Computer Science, Jagiellonian University, Kraków
}%
\author{Shilpa Mahato}
\affiliation{%
Department of Physics, Indian Institute of Technology Dhanbad
}%
\author{Rijul Sachdeva}
\affiliation{%
Jülich Supercomputing Center, Forschungszentrum Jülich
}
\affiliation{%
RWTH Aachen University
}%

\author{Debasmita Bhoumik}
\affiliation{%
Advanced Computing \& Microelectronics Unit, Indian Statistical Institute, Kolkata
}%

\author{Ritajit Majumdar\,\orcidlink{0000-0003-0730-0084}\,}
\email{ritajit\_r@isical.ac.in}
\affiliation{%
Advanced Computing \& Microelectronics Unit, Indian Statistical Institute, Kolkata
}%

\date{\today}

\begin{abstract}
Surface codes are quantum error correcting codes typically defined on 2D array of qubits. In this paper, a $[d_x,d_z]$ surface code design is being introduced, where $d_x (d_z)$ represents the distance of the code for bit (phase) error correction, motivated by the fact that the severity of bit flip and phase flip errors in the physical quantum system is asymmetric. We present pseudo-threshold and threshold values for the proposed surface code design for asymmetric error channels in the presence of various degrees of asymmetry of Pauli $\hat{X}$, $\hat{Y}$, and $\hat{Z}$ errors in a depolarization channel. We demonstrate that compared to symmetric surface codes, our asymmetric surface codes can provide almost double the pseudo-threshold rates while requiring less than half the number of physical qubits in the presence of increasing asymmetry in the error channel. Our results show that for low degree of asymmetry, it is advantageous to increase $d_x$ along with $d_z$. However, as the asymmetry of the channel increases, higher pseudo-threshold is obtained with increasing $d_z$ when $d_x$ is kept constant at a low value. Additionally, we also show that the advantage in the pseudo-threshold rates begins to saturate for any possible degree of asymmetry in the error channel as the surface code asymmetry is continued to increase.
\end{abstract}

\maketitle


\section{\label{asymm:Introduction}Introduction}

Quantum computers use certain quantum mechanical phenomena like superposition and entanglement to attain a substantial speedup over their conventional classical counterparts \cite{Harrow2017, Shor_1997, Aaronson2008}. Therefore, they are speculated to help solve certain problems that are intractable for even the most powerful classical supercomputers in the areas of drug discovery \cite{ZINNER20211680}, artificial intelligence \cite{Acampora2019}, material simulations \cite{Ma2020}, etc. However, the current generation of quantum hardware, generally referred to as Noisy Intermediate-Scale Quantum (NISQ) hardware \cite{Preskill2018}, has limited computational capabilities due to a small number of qubits, restrictive hardware connectivity, and poor qubit quality \cite{Preskill2018}. A significant effort is being put forward into combating these, and possibly developing feasible quantum error correction strategies for mitigating the effects of external noise on the qubits \cite{PhysRevX.10.011022, X49613} and making large-scale quantum computation viable \cite{ChaoNature2018, LitinskiQuantum2019}.

One possible approach is to use a collection of physical qubits to construct a logical qubit that would show more resilience against noise. Based on this approach, many quantum error correction codes (QECC) such as the Shor code \cite{Calderbank1996}, Steane code \cite{steane1996error}, etc. have been proposed in the past decades. However, these codes often suffer from the nearest-neighboring problem where overhead in the number of gates and error correction cycle is huge in executing quantum circuits on hardware with restrictive topologies due to interaction requirements amongst non-adjacent qubits \cite{majumdar2017method}. Topological codes were introduced to overcome this drawback \cite{Kitaev2003}. Surface code is one such topological QECC acting on a two-dimensional lattice of qubits with nearest-neighbour coupling. They have been shown to have a high tolerance for local errors, i.e., errors that can be corrected using just local operators acting on qubits placed in a two-dimensional grid \cite{Wang_2003}. Due to their structure they are scalable and at the same time have a high error threshold, of a value approaching 1\% \cite{Fowler_2009}.

Most of these codes were studied for the symmetric noise model, where each of Pauli $\hat{X}$ (bit flip), $\hat{Z}$ (phase flip), and $\hat{Y} = i\hat{Z}\hat{X}$ error occur with equal probability. However, with the physical implementations of quantum computers now being realized, it has been observed that many physical quantum channels are biased, i.e., the probability of one type of error is generally much higher than other types of errors \cite{PhysRevA.75.032345,1996}. Owing to this fact, the quantum error correction schemes must exploit this biasedness \cite{Tuckett2019}. Among the most recently studied quantum codes for this purpose are the surface codes \cite{Xu2019, Chamberland_2022_universal}, toric codes \cite{HansenArxiv2017}, skew cyclic codes \cite{IrwansyahAIP2019}, bosonic codes \cite{Chamberland_2022} etc. 

\textit{Major Contributions}: In this work, we propose a generalized and scalable $[d_x,d_z]$ surface code for the asymmetric depolarization noise model. Here $d_x$ and $d_z$ represent the distance of the code for correcting Pauli $\hat{X}$ (bit flip) and $\hat{Z}$ (phase flip) errors respectively. We demonstrate that it is straightforward to design the code from existing surface code for the symmetric noise model, thus making it easy to implement in near-term quantum devices. Using the MWPM decoder, we show that this code outperforms the existing surface code model in the presence of asymmetry in the noisy channel. Additionally, using numerical results, we demonstrate the relationship between asymmetry in the surface code and asymmetry in the error channel in relation to the pseudo-threshold and threshold values. We show that our proposed asymmetric surface codes can provide almost double the pseudo-threshold rates while requiring less than half the number of physical qubits compared to symmetric surface codes. For example, the asymmetric surface code $[3, 5]$ (Fig. \ref{fig:asymm-model-design}) achieves almost twice the pseudo-threshold rates than both $[5, 5]$ and $[7, 7]$ in a channel where the ratio of bit flip errors to phase flip errors is at most $1/10$ and requires $59.18\%$ and $70.10\%$ lesser physical qubits than them, respectively. Furthermore, since the design of surface codes has rotational symmetry, a rotation by $\pi/2$ will be sufficient if we ever have a channel that is biased towards the bit flip errors. Moreover, it is straightforward to modify this code using more qubits to attain even better performance in the presence of higher degree of asymmetry. There are other noise models studied in the literature, such as the amplitude damping, phase damping, erasure model which we do not consider in this study.

\textit{Structure}: Section \ref{asymm:SurfaceCode} briefly reviews the theory of surface code, the stabilizer formalism, and the working principle of Minimum Weight Perfect Matching (MWPM) based decoder. Section \ref{asymm:AsymmetricNoise} consists of a short introduction to the asymmetric noise model and a summary of existing Quantum Error Correction Codes (QECCs) for it. Section \ref{asymm:ProposedSurfaceCode} describes the structure of our proposed surface code, along with the logical errors that can occur in it and its scalability for different levels of asymmetry. Section \ref{asymm:Results} presents the decoding results, the performance of our proposed structure, and the comparison of the symmetric and the asymmetric surface codes. Finally, Section \ref{asymm:Conclusions} summarizes and concludes the discussion and the achieved results.

\section{\label{asymm:SurfaceCode}Surface Codes}
\begin{figure}[!tp]
    \centering
    \includegraphics[width=0.7\linewidth]{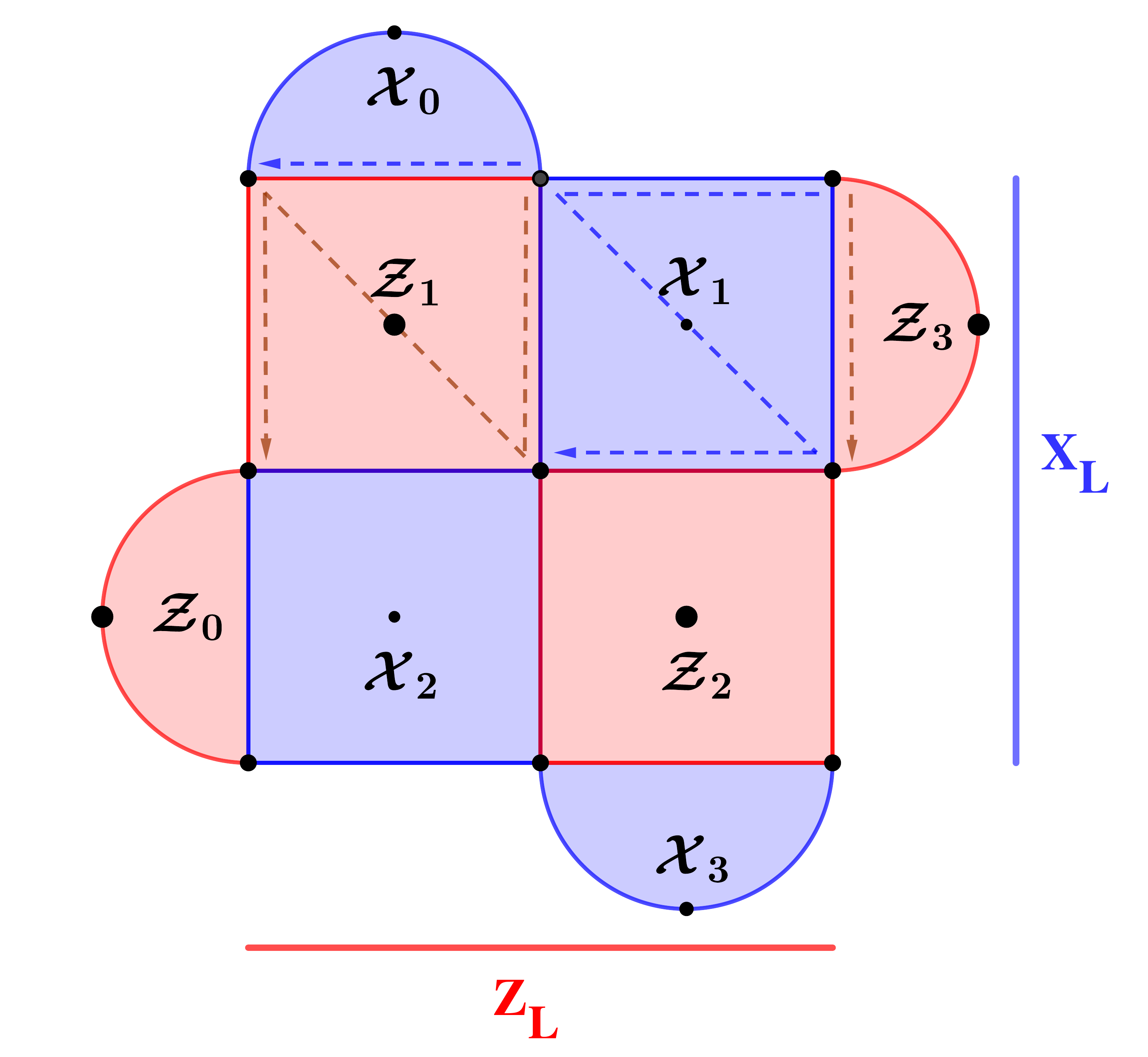}
    \caption{A distance 3 Surface Code}
    \label{fig:symm-model-design}
\end{figure}

Surface codes lie at the intersection of Topological codes and Stabilizer codes and are implemented on a two-dimensional lattice of physical qubits \cite{article}. They make use of the stabilizer formalism, where the stabilizer operators perform error detection and correction. These operators are abelian sub-groups of Pauli groups, and eigenvalues of their generators encode information regarding any possible occurrence of an error. Overall, the surface codes consist of two types of physical qubits: (i) data qubits, on which the actual quantum computation is performed, and (ii) ancilla qubits, on which syndrome measurement occurs for detecting errors. The latter consists of $X$ syndrome and $Z$ syndrome qubits, each of which interacts with the neighboring data qubits through parity checks, i.e., the value of the measurement, which is determined by the state of the connected data qubits. In Fig. \ref{fig:symm-model-design}, we show one of the smallest surface codes consisting of nine data qubits and eight ancilla qubits. Here, the stabilizers are represented by the square and semi-circular tiles, where ancilla qubits lie in the plaquette (or face of the lattice) and data qubits are on the vertices \cite{Varsamopoulos2018DesigningNN}, and their parity check operations can be represented as:
\begin{equation}
\begin{split}
    \mathcal{X}_0=&X_0X_1\quad \mathcal{X}_1=X_1X_2X_4X_5\\
    \mathcal{X}_2=&X_3X_4X_6X_7 \quad \mathcal{X}_3=X_7X_8\\
\end{split}
\end{equation}
\begin{equation}
\begin{split}
    \mathcal{Z}_0=&Z_3Z_6\quad \mathcal{Z}_1=Z_0Z_1Z_3Z_4\\
    \mathcal{Z}_2=&Z_4Z_5Z_7Z_8 \quad \mathcal{Z}_3=Z_2Z_5\\
\end{split}
\end{equation}

\begin{figure*}[htp]
    \centering
    \hfill
    \begin{subfigure}[b]{0.24\linewidth}
         \includegraphics[width=\linewidth]{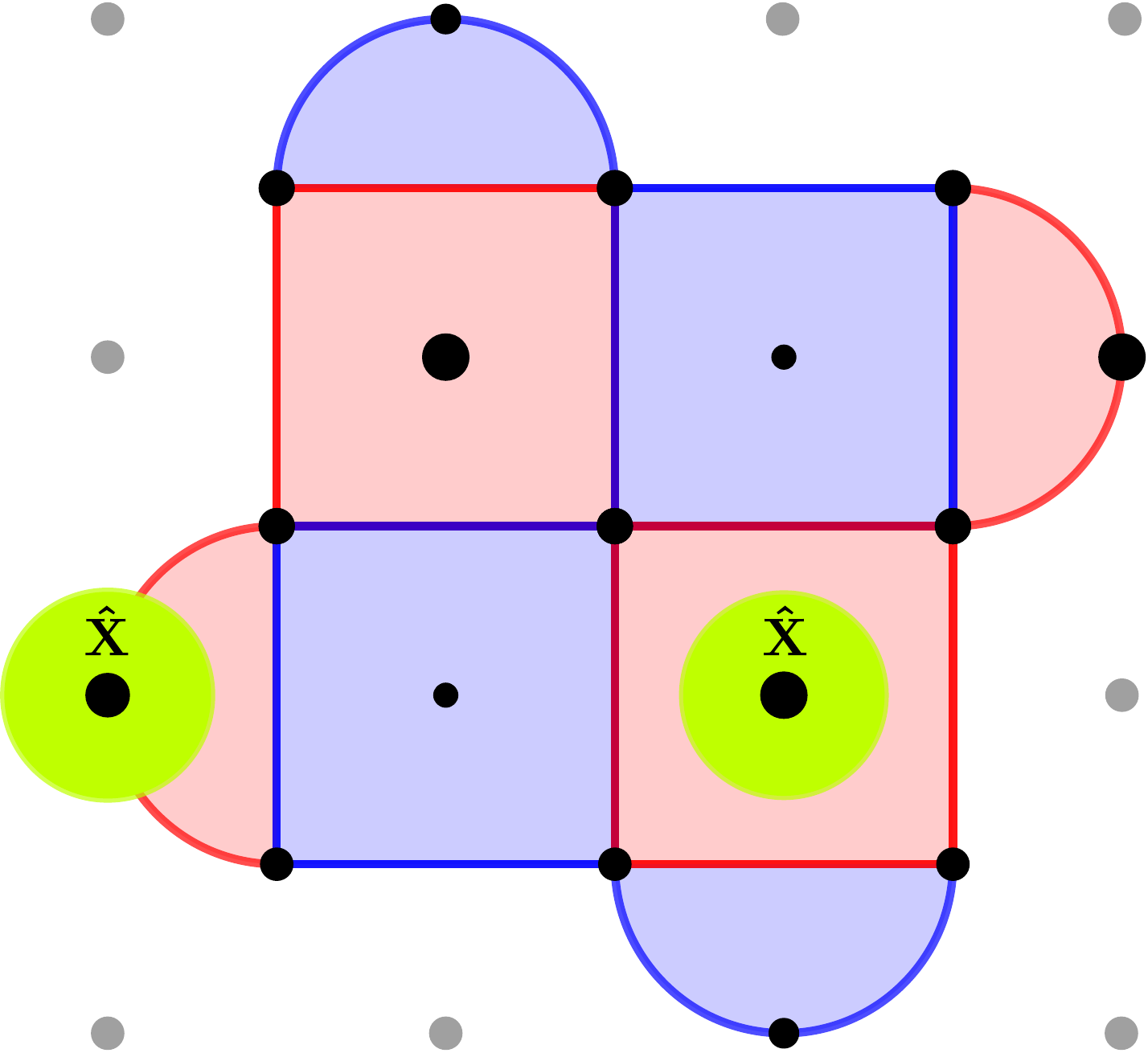}
        \caption{\label{fig:error-chain}}
    \end{subfigure}
    \hfill
    \begin{subfigure}[b]{0.20\linewidth}
        \includegraphics[width=\textwidth]{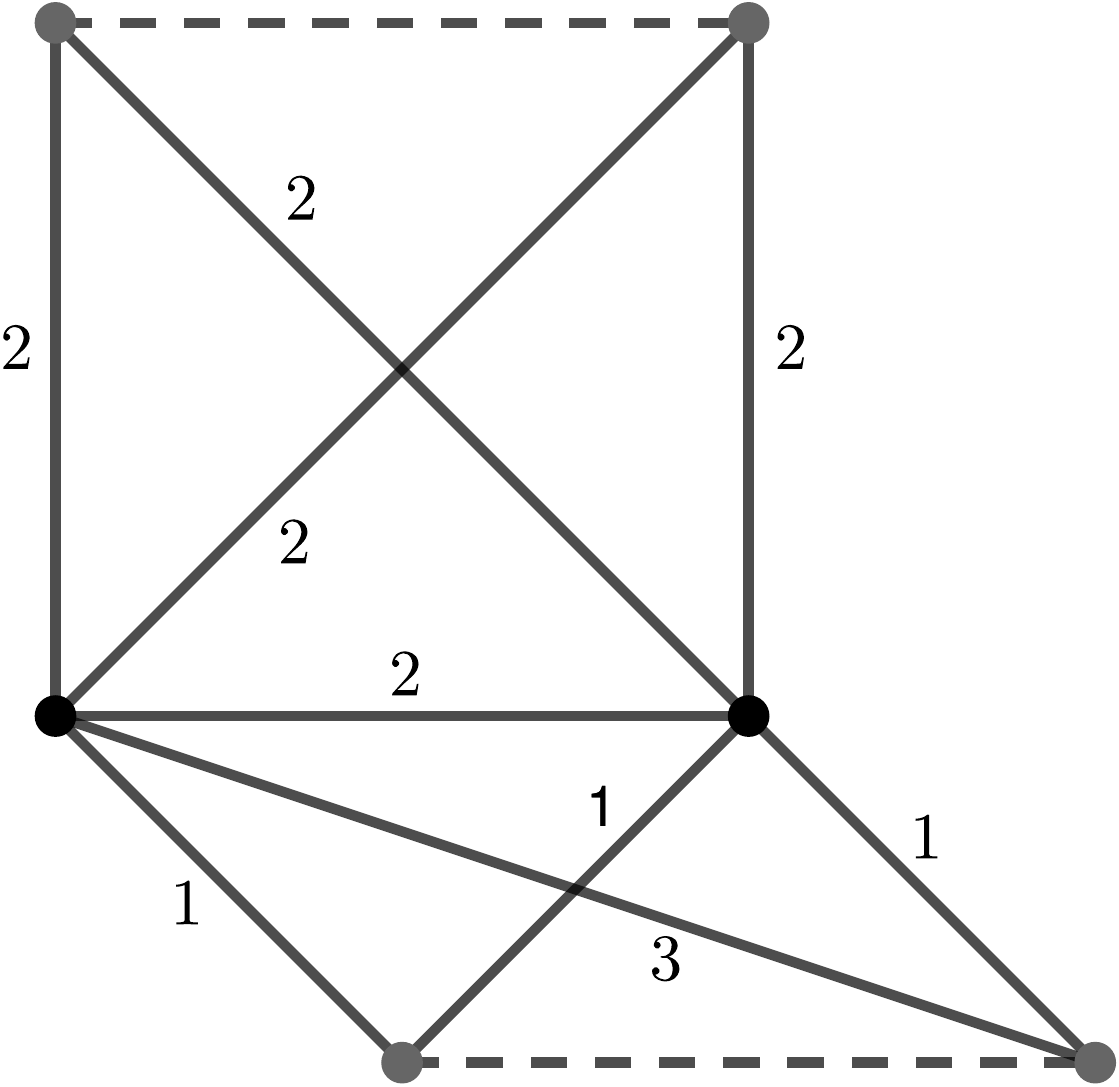}
        \caption{\label{fig:edge-graph}}
    \end{subfigure}
    \hfill
    \begin{subfigure}[b]{0.20\linewidth}
        \includegraphics[width=\textwidth]{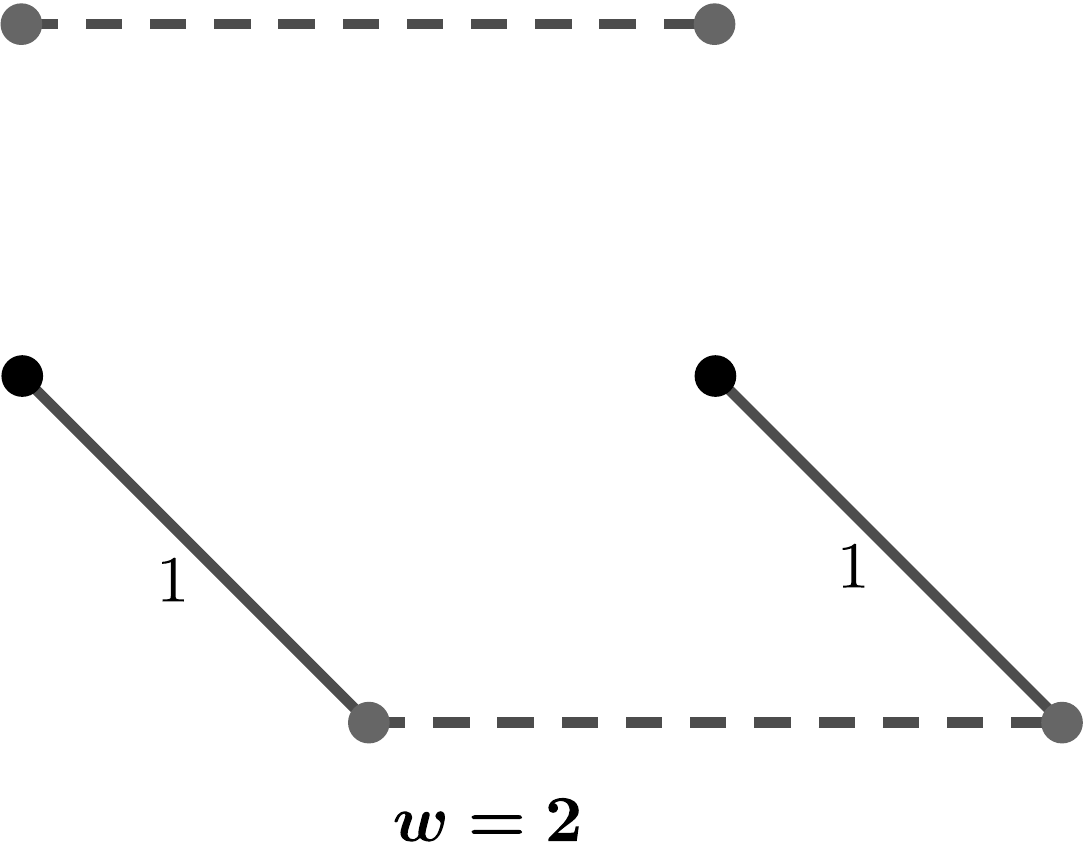}
        \caption{\label{fig:mwpm-surface}}
    \end{subfigure}
    \hfill
    \begin{subfigure}[b]{0.22\linewidth}
        \includegraphics[width=\textwidth]{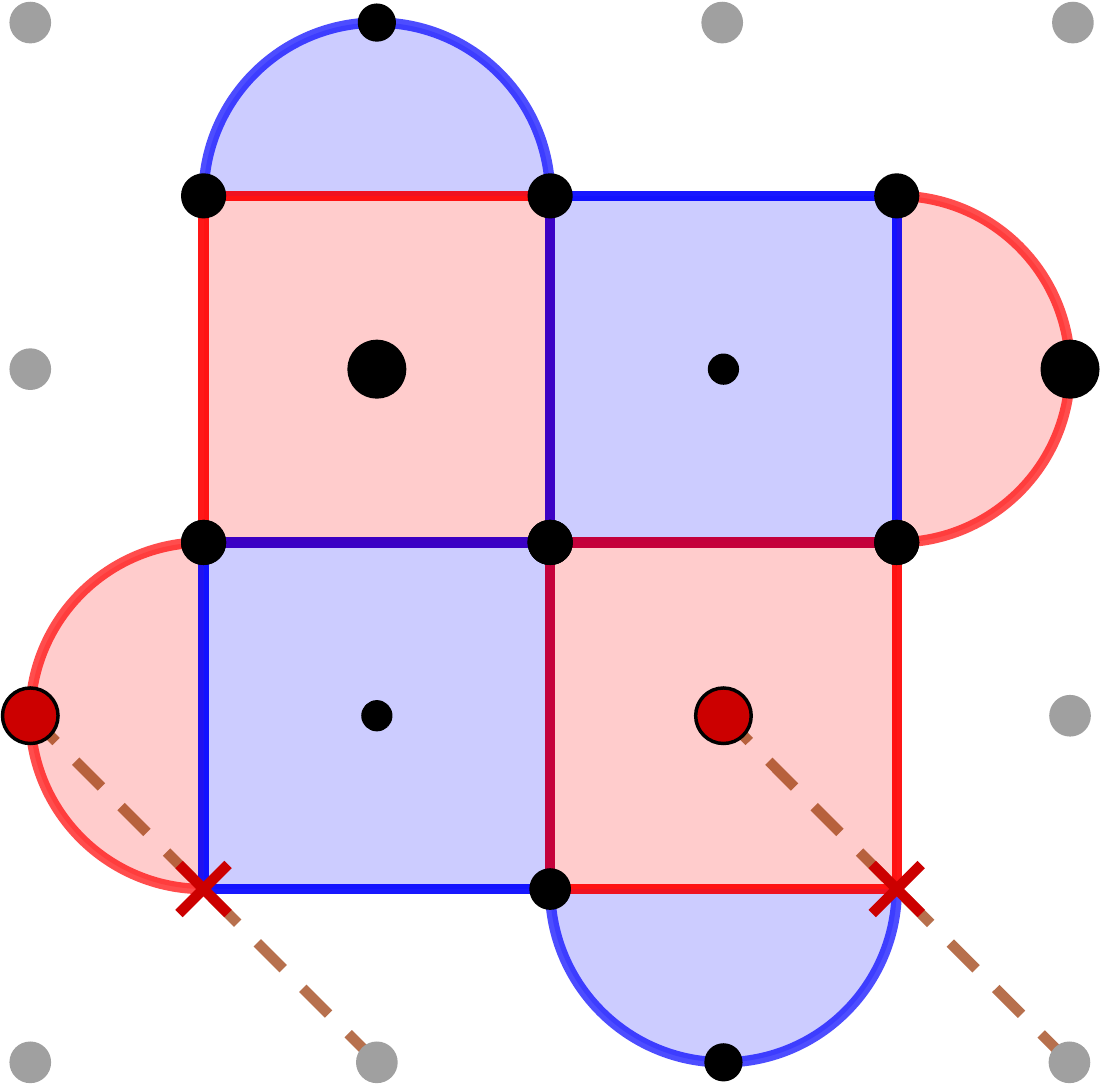}
        \caption{\label{fig:error-correction}}
    \end{subfigure}
    \hfill
    \caption{Error correction via the minimum weight matching (MWPM) algorithm. (a) Error syndromes are highlighted in green and labeled with the type of error that puts them in $-1$ state. (b) Error syndromes (highlighted black nodes) and dummy boundary nodes (gray) are mapped to a weighted graph. The dotted lines here represent the boundaries. (c) The goal is to find a subgraph such that each vertex (error syndrome) with either another vertex or a boundary such that the total weight of matched edges is minimal. Here we represent one such possible subgraph with weight $w=2$. (d) Using the subgraph, data qubits on which the error occurred are identified (marked with red crosses). They are corrected, and all error syndromes (displayed in red) are restored to $+1$ state again.}
    \label{fig:surface-code-error-correction-example}
\end{figure*}

In each surface code cycle, $X$ and $Z$ syndrome qubits are initialized in their ground state and entangled with the neighboring data qubits using CNOT gates. At the end of the surface code cycle, an Error Syndrome Measurement (ESM) is performed on all the syndrome qubits. These measurements measure the $\hat{X}$ or $\hat{Z}$ stabilizers, i.e., their outcome can predict the presence of errors without perturbing the system. If the parity checks for the current surface code cycle are the same as the previous one, we conclude that the state of the data qubits involved in the parity checks has not changed due to any erroneous operation. However, if the state of an odd number of data qubits involved in a parity-check is changed, the parity-check returns a different value ($0\leftrightarrow1$), triggering a \textit{detection} event. Hence, the state of the system $\ket{\psi}$ that it is prepared into at the end of the first surface code cycle is maintained as long as errors don't occur, and this state is known as the \textit{quiescent state}. The altered values are passed to a decoder, which combines them to identify the possible location of the errors, i.e., the indices of the data qubits. 

In general, there is no decoding algorithm that provably works best for all scenario. But, the most widely used one is the MWPM \cite{Fowler2015MinimumWP}, in which a weighted graph is generated where the syndromes serve as nodes, and dummy nodes as boundary nodes \cite{Hadfield2015ErrorRT}. Each node can connect to either of the four boundary nodes by hopping over one or more physical qubit(s) forming an edge. A weight is associated with this edge, depending upon the number of physical qubits hopped. The edges with the minimum weight are considered, and the physical qubits hopped over are marked as erroneous and are corrected accordingly (Fig. \ref{fig:surface-code-error-correction-example}).

Other than this, look-up table \cite{varsamopoulos2017decoding} and machine learning (ML) based decoders have been proposed for surface codes as well \cite{bhoumik2021efficient,Varsamopoulos2018DesigningNN,sweke2018reinforcement}. Even though ML decoders have been observed to outperform MWPM decoders for symmetric noise model, it is not known whether the same ML model with the same parameters will provide the best performance for an asymmetric model with any degree of asymmetry (since the structure becomes more and more rectangular). This would be an interesting topic for future work. Since MWPM decoder carries over naturaly from square lattice to a rectangular one, we stick with the MWPM decoder for this study.

As illustrated in Section \ref{asymm:logical_op}, all of the physical qubits present in the surface can be encoded as one or more logical qubits that are expected to be more resilient against errors. These logical qubits can be manipulated by the logical operators $\hat{X}_L$ and $\hat{Z}_L$, which are defined as a chain of Pauli $\hat{X}$ and Pauli $\hat{Z}$ operators spanning between two vertical and horizontal boundaries respectively. The length of these operators is $\geq d$, where $d$ is defined as the distance of the surface code \cite{Varsamopoulos2018DesigningNN}. The number of errors (t) that a surface code can successfully correct is also determined by its distance  $d$ as per the following equation:
\begin{equation}
t = \bigg\lfloor\frac{d-1}{2}\bigg\rfloor
\end{equation}
Thus, a $d=3$ surface code (Fig.~\ref{fig:symm-model-design}) will be successfully able to correct up to a single logical $\hat{X}$ and logical $\hat{Z}$ error.

At the moment, our simulation considers errors only on data qubits.The study of the performance where both data and measure qubits can be erroneous is reserved as a future extension of this research. The decoding performance is usually quantified by two metrics: threshold and pseudo-threshold. The \textit{threshold} value helps to characterize the decoding performance of a surface code design and is defined as the point of intersection of logical error rate curves for different code distances. Beyond this, increasing the distance of the code leads to higher probability of logical error. The \textit{pseudo-threshold} value determines the highest probability of physical error below which error correction leads to lower logical error probability. It is defined as the point of intersection of logical error rate curves for a given code distance with the curve for which the physical error rate is equal to the logical error rate.

\section{\label{asymm:AsymmetricNoise}The Asymmetric Noise Model}

Quantum error-correcting codes serve to correct errors resulting from noise inherent in modern quantum computers. Studying the nature of the physical noise is crucial for creating QECCs with low levels of redundancy. The noise is typically asymmetric in modern physical quantum devices, and phase errors are much more probable than bit flips. It's possible to characterize the level of noise by two parameters, relaxation time and dephasing time. While dephasing only results in phase flip errors, relaxation results in both phase flip errors and bit flips \cite{PhysRevA.75.032345}. This leads to a large asymmetry, where phase flips occur much more frequently than bit flips.

There exist several QECCs that exploit this asymmetry of errors. One of the most popular approaches is to concatenate two QECC codes. The simplest codes of this nature are presented in \cite{PhysRevA.78.052331}, where repetition code is concatenated with CSS code, and in \cite{doi:10.1142/S1230161211000029}, where a concatenation of repetition code and error avoiding code is presented. In \cite{Stephens_2008}, symmetric code is combined with an asymmetric one, and fault-tolerant circuits are needed to switch between the symmetric and asymmetric encodings. Both \cite{PhysRevA.75.032345} and \cite{doi:10.1098/rspa.2008.0439} analyze a CSS construction, which uses a classical LDPC code for the more common phase errors, and a classical Bose–Chaudhuri–Hocquenghem (BCH) code for bit flips. In \cite{evans2007error} the authors use a different approach, where a higher frequency is used for syndrome measurements of $X$-only generators as compared to the $Z$-only generators. Some approaches have also focused on creating the shortest possible QECC for the asymmetric noise model. For example, in \cite{10.1007/978-3-030-50433-5_49}, the CSS-based codes are constructed by a specific syndrome assignment, and in \cite{Jackson_2016}, a random search of codes of up to length $9$ is performed in the hope of finding ones with desirable properties.

Closest to our work are the approaches that utilize surface code to tackle the asymmetric noise model. In \cite{Xu_2019}, a variant of surface code is introduced, which is concatenated with a two-qubit phase detection code. A decoding scheme based on toric code was used in \cite{Nickerson_2019} and tested using the MWPM decoder \cite{Fowler2015MinimumWP}, which we also perform on our design. In \cite{Tuckett2019, Tuckett2020} authors demonstrate the advantages of using a rectangular form of traditional surface-code, albeit with a different decoding scheme that exploits the symmetries of its syndrome. Finally, in \cite{BonillaAtaides2021} and \cite{Darmawan2021}, the performance of XZZX surface codes has been studied for asymmetric noise and circuit noise model with Kerr-cat qubits, respectively.

This work aims to design a surface code for an asymmetric noise model, increasing the error-correction thresholds while still being easy to implement, and test it using an MWPM decoder.

\section{\label{asymm:ProposedSurfaceCode}Proposed Surface Code Design}

In the past, surface codes have generally been designed to protect the qubits against equiprobable Pauli errors, i.e., the bit flip $X$, phase flip $Z$ and combined bit-phase flip $Y$ errors. Inherent to their construction, their error correction capabilities depend upon their distance $d$, i.e., a surface code of distance $d$ can correct up to $t = \lfloor(d-1)/2\rfloor$ Pauli $\hat{X}$ and $\hat{Z}$ errors. Usually, it is assumed that the surface codes are symmetric, i.e., $d_x=d_z=d$. However, as explained in Section \ref{asymm:AsymmetricNoise}, Pauli errors don't occur symmetrically in all quantum channels. Instead, the phase flips errors $Z$ happen with different severity than the bit flip errors $X$ for real quantum systems, majorly due to the complexity of the noise processes. Therefore, it feels natural to include this asymmetry in the design of surface codes as well. We propose to do this by introducing asymmetry in the distance of the surface codes, i.e., we represent an asymmetric surface code by $[d_x, d_z]$. This way, such a surface code can correct up to $t_x = \lfloor(d_x-1)/2\rfloor$ bit flip errors and $t_z = \lfloor(d_z-1)/2\rfloor$ phase flip errors, and the symmetric surface codes become a special case of asymmetric surface codes with $d_x=d_z$.

\begin{figure}[t]
    \centering
    \includegraphics[width=\linewidth]{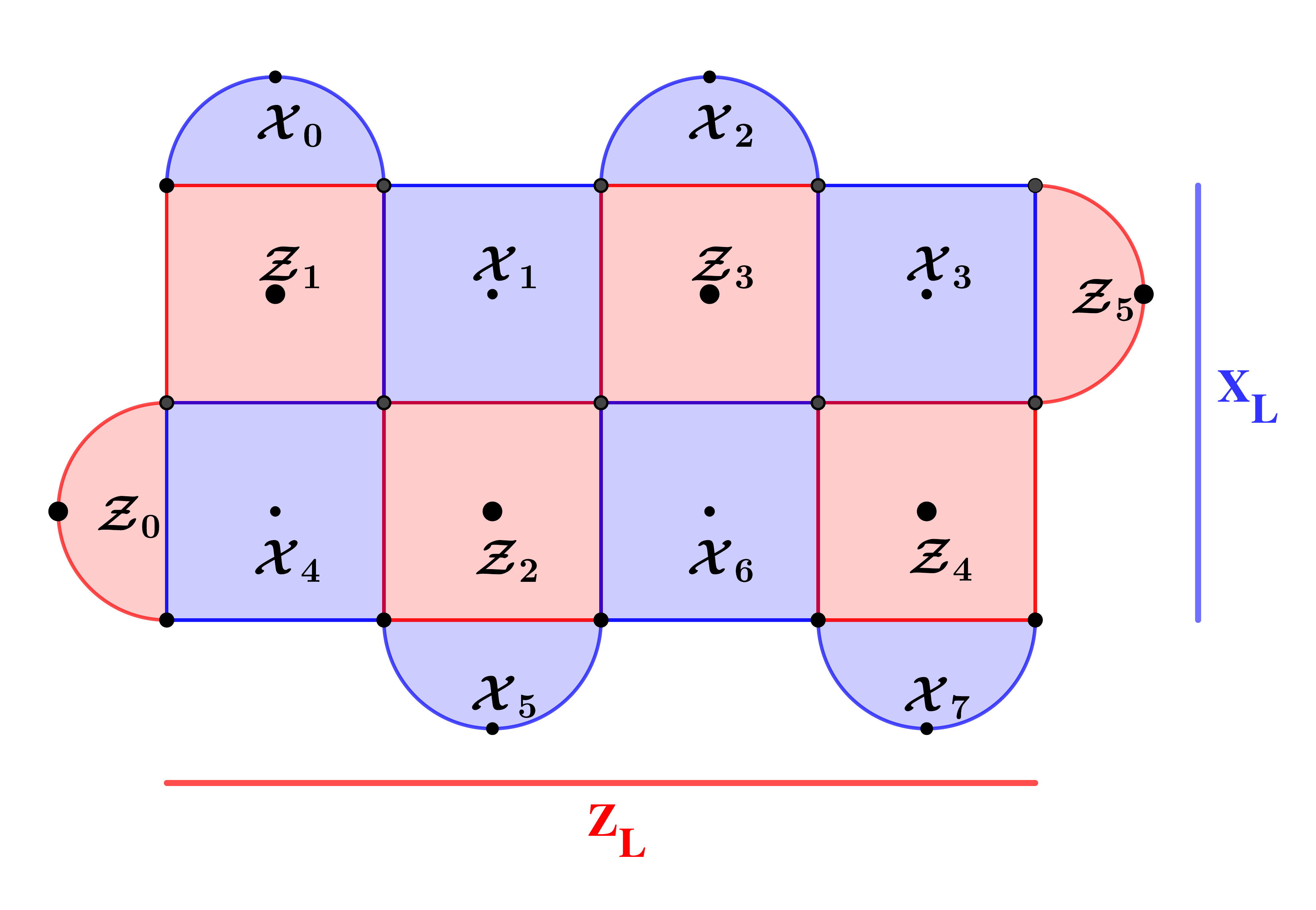}
    \caption{Asymmetric surface code having distance $[3, 5]$. $Z$ and $X$ measure qubits are shown in red (indexed $ \mathcal{Z}_0- \mathcal{Z}_5$), and blue (indexed $ \mathcal{X}_0- \mathcal{X}_7$) respectively.}
    \label{fig:asymm-model-design}
\end{figure}

Now depending upon the asymmetry in the channel, one can choose whether $d_x>d_z$ should be true or $d_x < d_z$. Since we assume that bit flip errors are less prevalent than phase flip errors, we choose $d_x<d_z$. A symmetric surface code of distance $d$ is a $d \times d$ square lattice. Since logical Z (discussed in detail in next subsection) is, in general, defined as a horizontal operator, our asymmetric surface code will be a $d_z \times d_x$ lattice. An example design of the asymmetric surface code $[3, 5]$ is illustrated in Fig.~\ref{fig:asymm-model-design}. The position of data and ancilla qubits remain the same as in symmetric surface code, making this straight-forward to design. Our example $[3,5]$ asymmetric surface code has $8\ X$ stabilizers and $6\ Z$ stabilizers which can correct up to one bit flip and two phase flip errors respectively. 

\subsection{Logical Operators\label{asymm:logical_op}}

For any given surface code $[d_x, d_z]$, we have $d_x d_z$ data qubits and $d_z d_x-1$ measurement qubits, which gives $2^{d_x d_z}$ degrees of freedom and $2^{(d_x d_z - 1)}$ degrees of constraints. The two unconstrained degrees of freedom allow us to consider the entire lattice as one logical qubit that can be manipulated by logical operators $\hat{X}_{L}$ and $\hat{Z}_{L}$. We build these logical operators by looking at multi-qubit operator products that commute with the stabilizers and connect the opposite boundaries. Similar to the case of symmetric surface codes, the logical operators $\hat{X}_{L}$ and $\hat{Z}_{L}$ are defined by $\hat{X}$ and $\hat{Z}$ operations on the data qubits in each column and row, respectively, where $||\hat{X}|| = d_x$ and $||\hat{Z}|| = d_z$. These are represented in the Fig. \ref{fig:asymm-model-design}.

\begin{figure}[t]
    \centering
    \includegraphics[width=\linewidth]{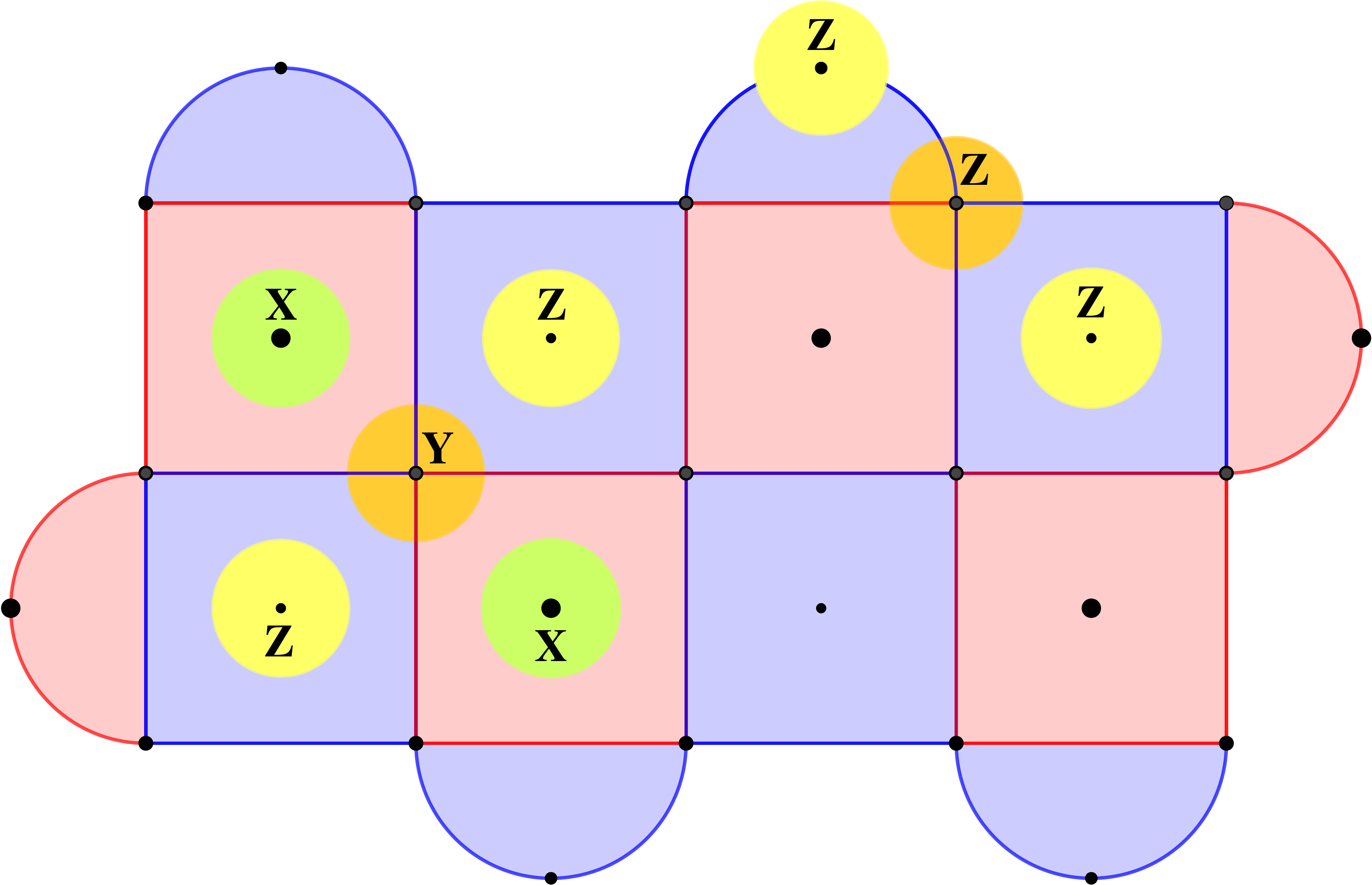}
    \caption{The working of asymmetric surface code $[3, 5]$. $Z$ and $X$ measure qubits shown in red and blue, respectively, are measured in each surface code cycle. $\hat{X}$ ($\hat{Z}$) errors on an odd number of data qubits involved in the $Z$ ($X$) syndrome measurements will result in $-1$ eigenvalues triggering a detection event shown in green (yellow). Here, the Pauli $\hat{Y}=i\hat{Z}{X}$ error on data qubit indexed $6$ is captured by the syndromes for both $Z$ and $X$ stabilizers. Whereas the Pauli $\hat{Z}$ error on data qubit indexed $3$ affects the syndromes for only $X$ stabilizers. The $X$ ($Z$) syndrome measurement outcomes are represented by a binary vector where all the syndromes involved in the detection events have a value of $1$. These binary vectors are sent to the MWPM decoder, which combines them to identify the possible indices of data qubits that suffered the errors.}
    \label{fig:asymm-model-working}
\end{figure}

\begin{equation}
\begin{split}
    \hat{X}_{L} \in\ & \{\hat{X}_{0}\hat{X}_{5}\hat{X}_{10},\  \hat{X}_{1}\hat{X}_{6}\hat{X}_{11},\  \hat{X}_{2}\hat{X}_{7}\hat{X}_{12},\\
    &\hat{X}_{3}\hat{X}_{8}\hat{X}_{13},\  \hat{X}_{4}\hat{X}_{9}\hat{X}_{14}\}
\end{split}
\end{equation}

\begin{equation}
\begin{split}
    \hat{Z}_{L} \in\ & \{\hat{Z}_{0}\hat{Z}_{1}\hat{Z}_{2}\hat{Z}_{3}\hat{Z}_{4},\  \hat{Z}_{5}\hat{Z}_{6}\hat{Z}_{7}\hat{Z}_{8}\hat{Z}_{9},\\
    &\hat{Z}_{10}\hat{Z}_{11}\hat{Z}_{12}\hat{Z}_{13}\hat{Z}_{14}\}
\end{split}
\end{equation}

In addition to these, one can choose some other multi-qubit operator that satisfies the above conditions to build other logical operators $\hat{X}_{L}^{\prime}$ and $\hat{Z}_{L}^{\prime}$. However, these operators will be linearly dependent on the previously defined logical operators $\hat{X}_{L}$  and $\hat{Z}_{L}$ respectively. For example, consider the following case
\begin{equation}
    \begin{split}
        \hat{X}_{L}^{\prime} =& \hat{X}_{0}\hat{X}_{6}\hat{X}_{12}\\
        =& \hat{X}_{1}\hat{X}_{2}\hat{X}_{5}\hat{X}_{7}\hat{X}_{10}\hat{X}_{11}(\hat{X}_{0}\hat{X}_{5}\hat{X}_{10})(\hat{X}_{1}\hat{X}_{6}\\
        &\hat{X}_{11})(\hat{X}_{2}\hat{X}_{7}\hat{X}_{12})\\
        =& (\hat{X}_{1}\hat{X}_{2}\hat{X}_{7} \hat{X}_{6})(\hat{X}_{6}\hat{X}_{5}\hat{X}_{10}\hat{X}_{11})\hat{X}_{L}^0\hat{X}_{L}^1\hat{X}_{L}^2
    \end{split}
\end{equation}
In this case, $X^{\prime}_{L}$ turns out to be a few of $\hat{X}_{L}^{i}$ operators multiplied by couple of operator products representing stabilizers $X_1$ and $X_4$, which are stabilized to a $\pm1$ eigenvalue by the surface code. Using this fact, we can show that the action of  $X^{\prime}_{L}$ on the quiescent state $\ket{\psi}$ will be
\begin{equation}
    \begin{split}
        \hat{X}_{L}^{\prime}\ket{\psi} &= \hat{X}_{1,2,7,6}\hat{X}_{5,6,10,11}\hat{X}_{L}^0\hat{X}_{L}^1\hat{X}_{L}^2\ket{\psi}= \pm \ket{\psi_L}
    \end{split}
\end{equation}

This will hold true for any $X^{\prime}_{L}$ (or $Z^{\prime}_{L}$) that can be written as a product of stabilizers times $\hat{X}_L$ (or $\hat{Z}_{L}$) operators. Therefore, instead of looking at the complete, exhaustive list of possible logical operators, we can just focus on the logical operators $\hat{X}_{L}$  and $\hat{Z}_{L}$.

\subsection{Logical Errors}
Now, having defined logical operators, we can look at logical errors in the asymmetric surface code. Usually, a Pauli error on a single data qubit is more likely, but sometimes Pauli errors can occur on two or more data qubits together, creating error chains. As already explained in section \ref{asymm:SurfaceCode}, these are the physical errors as they occur on the data (physical) qubits. These can be identified by using the MWPM algorithm \cite{Fowler2015MinimumWP}, and then corrected manually with the help of some control software. For example, in Fig. \ref{fig:asymm-model-working}, we show an error chain $\hat{Y}_6\hat{Z}_3$ acting on the data qubits indexed $6$ and $3$. The error $\hat{Y}_6$ would trigger the stabilizers $Z_1$, $Z_2$, $X_1$ and $X_4$, whereas the $\hat{Z}_3$ would trigger the stabilizers $X_2$ and $X_3$. The syndrome for X stabilizers would be $[0, 1, 1, 1, 0, 0, 0, 0]$ and for Z stabilizers would be $[0, 1, 1, 1, 0, 0]$. These are used by MWPM for decoding and identifying which data qubits possibly suffered the errors.

However, there are times when MWPM misidentifies the errors, especially when the error chain formed is not sparse. In such cases, there's a possibility of errors persisting even after the correction. These errors are more dangerous when they are logical errors, i.e., they change the state of the logical qubit made by the surface code. It is only those errors, which anti-commute with at least one of the $\hat{X}$ (or $\hat{Z}$) logical operators forming the $Z$ (or $X$) logical errors. Therefore, these errors remain undetected, and affect the computation henceforth since the logical state has changed. For example, if in the previous example, the error chain would have been $\hat{Z}_2 \hat{Z}_4 \hat{X}_6 \hat{Z}_{10}$ instead of $\hat{Y}_6 \hat{Z}_3$, the syndromes generated (and hence the decoding result) would still have been the same, leading to an overall error of $\hat{Z}_2 \hat{Z}_4 \hat{Z}_3 \hat{Y}_6 \hat{Z}_{10}$. This is a logical $Z$  error because it anti-commutes with more than one $\hat{X}_{L}$ operators such as $\hat{X}_{0}\hat{X}_{5}\hat{X}_{10}$ or $\hat{X}_{2}\hat{X}_{7}\hat{X}_{12}$. Therefore, comparing how many such logical errors occur from a given number of decoding trials provides us with estimates of the asymmetric surface code's tolerance against errors, which we present in the next section. 

\begin{figure}[!tp]
    \centering
    \begin{minipage}{0.05\linewidth}
    (a)
    \end{minipage}%
    \begin{minipage}{0.95\linewidth}
    \begin{subfigure}{\linewidth}
        \includegraphics[width=\linewidth]{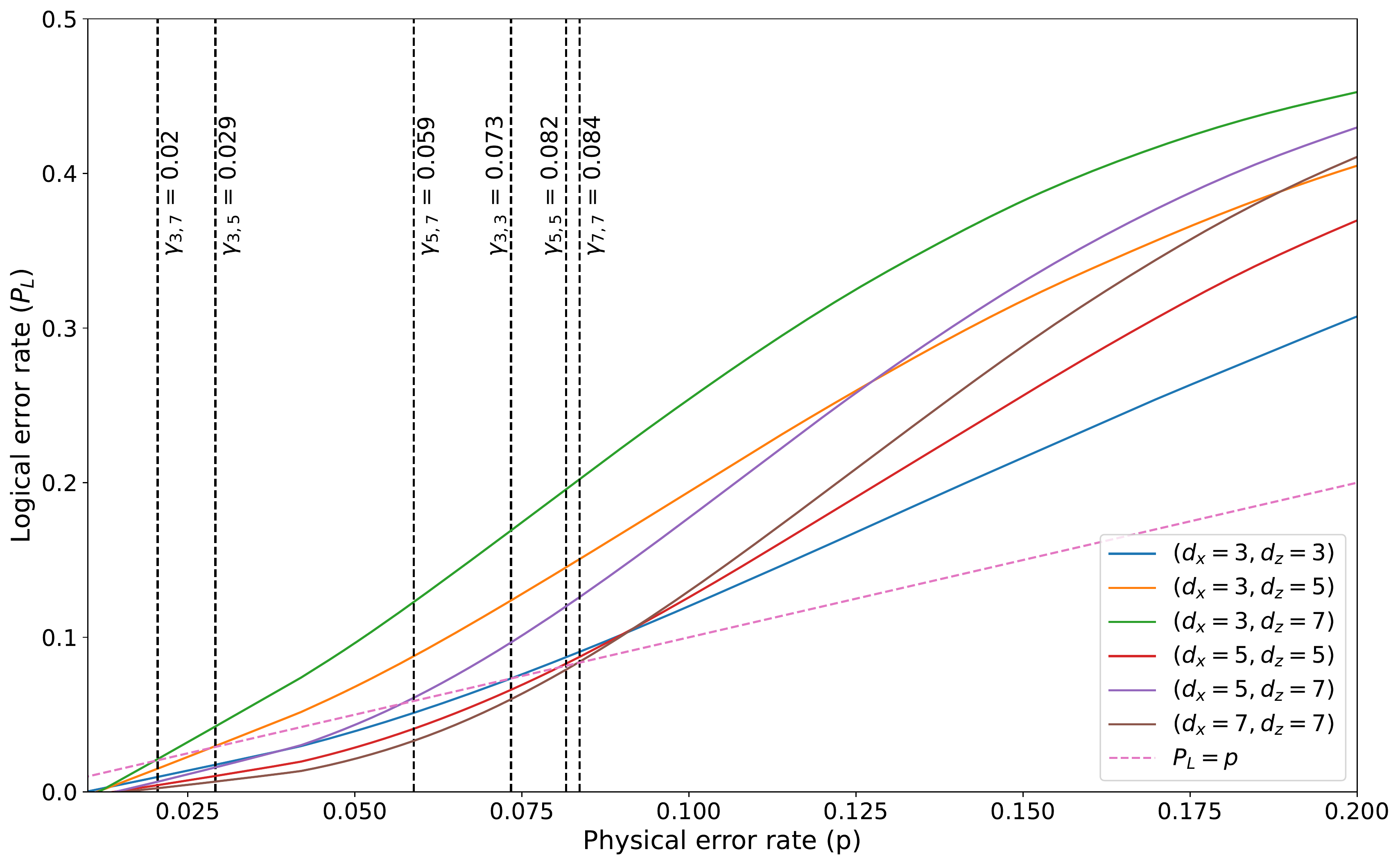}
    \end{subfigure}
    \end{minipage}\\
    \begin{minipage}{0.05\linewidth}
    (b)
    \end{minipage}%
  \begin{minipage}{0.95\linewidth}
    \begin{subfigure}{\linewidth}
        \includegraphics[width=\linewidth]{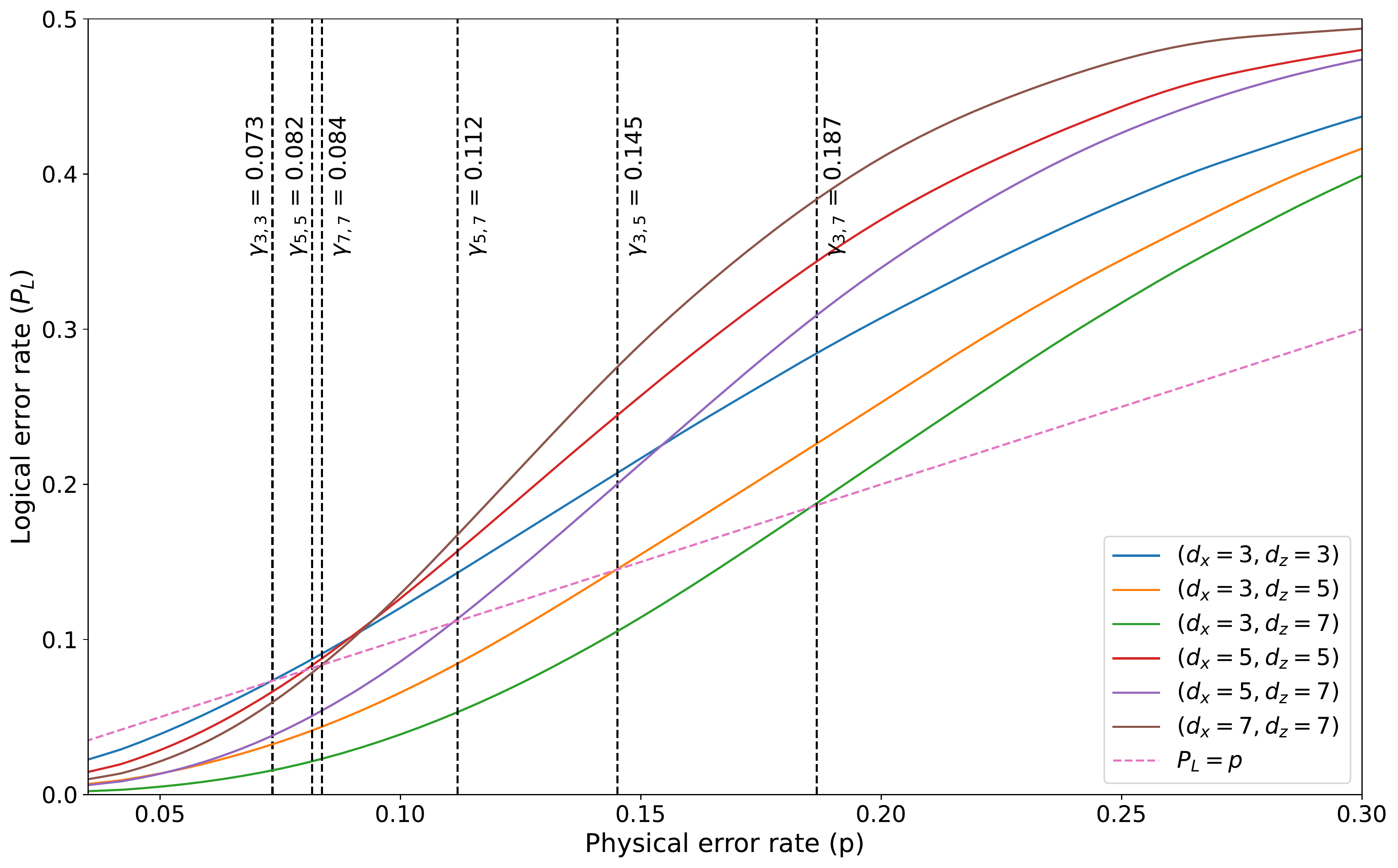}
    \end{subfigure}
    \end{minipage}
    \caption{Performance of asymmetric surface codes,\newline\{[3, 5], [3, 7], [5, 7]\}, in comparison to symmetric surface codes, \{[3, 3], [5, 5], [7, 7]\}, for correcting (a) bit flip errors and (b) phase flip errors. Pseduo-thresholds for each surface code $(d_x, d_z)$ is represented by $\gamma_{d_x, d_z}$.\label{fig:parity-phase}}
    \label{fig:pseudo}
\end{figure}

\section{\label{asymm:Results}Results}

In this section, we analyze the performance of our proposed surface code design for correcting errors with different levels of asymmetry in (i) code distance pair $[d_x, d_z]$, (ii) Pauli errors in the depolarizing noise channel, and (iii) a combination of both of them. 

As mentioned before, we use pseudo-threshold values to compare the decoding performance of two surface codes with different code distance pairs $[d_{x}, d_z^{i}]$ and $[d_{x}, d_z^{j}]$. This value is estimated for a given surface code by performing a simulation to calculate logical error rates $P_{L}$ for a range of physical error rates $p$, where the former is defined for any given value of the latter as the ratio of total logical errors accumulated in all error correction cycles to the number of net error correction cycles performed. Each of the simulations was averaged over $50,000$ repetitions, with the range of physical error $\kappa$ varying between $\kappa \in [10^{-4}, 5.5 \times 10^{-2}]$, which corresponded to the physical error rate of $p = 1-(1-\kappa)^8$ for the complete execution of a surface code cycle. Here, the factor of $8$ arises from the fact that each cycle consists of eight steps as shown in \cite{article}. 

Additionally, for decoding, we have used the PyMatching library \cite{higgott2021pymatching}, which implements the MWPM algorithm \cite{Fowler2015MinimumWP} to calculate the data qubits that need correction from a provided error syndrome.

\subsection{\label{asymm:sub:ParityPhase}Performance on Bit Flip (X) and Phase Flip (Z) Errors}

In the first set of experiments, we compared the performance of proposed asymmetric surface codes $[d_x, d_z]$, where $d_x, d_z \in \{3, 5, 7\}$ and $d_x<d_z$, against their corresponding symmetric counterparts $[d, d]$, where $d \in \{3, 5, 7\}$, for correcting bit flip (X) and phase flip (Z) errors, independently. To do this, we tested them for error channels with only one kind of error, either Pauli $\hat{X}$ or $\hat{Z}$, and not both. From the results presented in Fig. \ref{fig:parity-phase}, it is evident that there is an overall improvement in the performance of correcting phase errors as the asymmetry in the surface codes $[d_x, d_z]$ is increased. This can also be inferred from their higher pseudo-threshold values than their symmetric variants $[d_x, d_x]$ and $[d_z, d_z]$. For example, [3, 5] and [3, 7] have much better pseudo-threshold values than [3, 3], [5, 5] and [3, 3], [7, 7], respectively. 

However, at the same time, we also notice a considerable decrease in the performance of correcting bit flip errors for asymmetric surface codes. We attribute this decrease to an insufficient increase in the number of Z stabilizers in comparison to the rise in the number of possible combinations of $X_{L}$, due to an increase in the number of qubits ($d_x \times d_x \rightarrow d_x \times d_z$; $d_z > d_x$). For example, in the case of [3, 5], the increase in possible combinations of $X_{L}$ is from ${}^{9}C_3$ to ${}^{15}C_3$, while Z stabilizers increment from $4$ to only $6$. Similar logic can also explain the intermediate performance of $[5, 7]$ in both cases compared to the rest of the codes.

\begin{table}[t]
\centering
\caption{Variation in pseudo-threshold ($\gamma_{d_x, d_z}$) values with respect to step-wise change in the asymmetry of depolarizing noise.}
\label{table:symm-asymm-noise}
\begin{tabular}{|c|c|c|c|c|c|c|} 
\hline
\textbf{Asymmetry} & \multicolumn{6}{c|}{\textbf{Pseudo-thresholds}}   \\ 
\cline{2-7}
$(\Delta)$         & $\gamma_{3,3}$ & $\gamma_{3,5}$ & $\gamma_{3,7}$ & $\gamma_{5,5}$ & $\gamma_{5,7}$ & $\gamma_{7,7}$\\ 
\hline
1                  & 0.082          & 0.073          & 0.065          & 0.102          & 0.096          & 0.110\\ 
\hline
2                  & 0.094          & 0.108          & 0.096          & 0.110          & 0.122          & 0.118\\ 
\hline
3                  & 0.094          & 0.128          & 0.128          & 0.110          & 0.132          & 0.116\\ 
\hline
4                  & 0.096          & 0.137          & 0.151          & 0.108          & 0.136          & 0.112\\ 
\hline
5                  & 0.092          & 0.145          & 0.167          & 0.104          & 0.138          & 0.108\\ 
\hline
6                  & 0.094          & 0.151          & 0.176          & 0.102          & 0.136          & 0.104\\ 
\hline
7                  & 0.092          & 0.153          & 0.184          & 0.100          & 0.134          & 0.102\\ 
\hline
8                  & 0.09           & 0.155          & 0.187          & 0.098          & 0.132          & 0.100\\ 
\hline
9                  & 0.089          & 0.157          & 0.190          & 0.096          & 0.130          & 0.098\\ 
\hline
10                 & 0.088          & 0.155          & 0.192          & 0.095          & 0.128          & 0.098\\
\hline
\end{tabular}
\end{table}

\begin{figure*}[!htp]
    \centering
    \includegraphics[width=\linewidth]{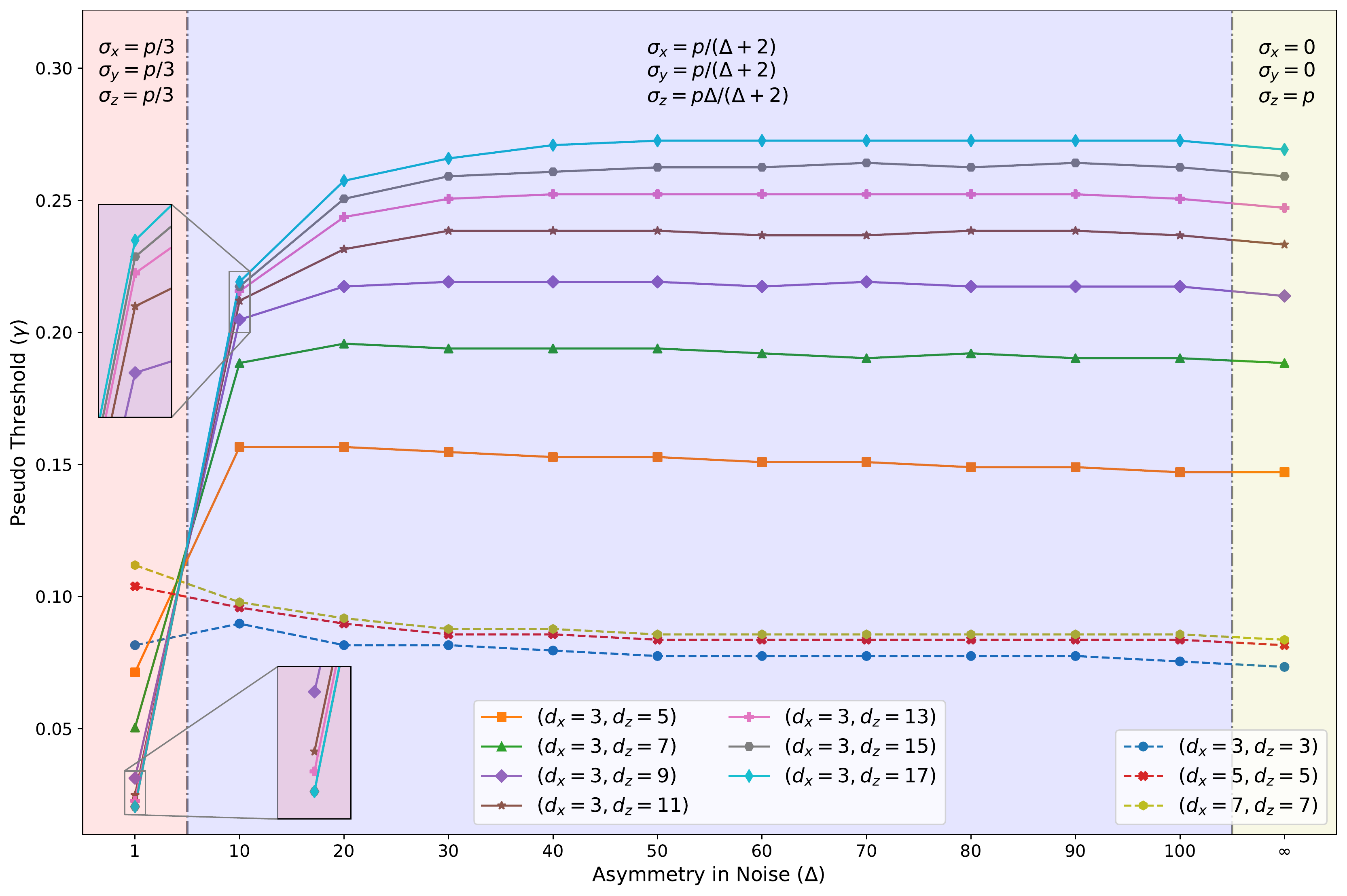}
    \caption{Analyzing change in pseudo-threshold values for different asymmetric and symmetric surface codes with the asymmetry in the noise. Here, we consider three regions of asymmetry demarcated by different values of $\Delta$ - (i) symmetric: $\Delta = 1$ (red), (ii) intermediate asymmetry: $\Delta \in [10, 100]$ (blue), and (iii) strong asymmetry: $\Delta \rightarrow \infty$ (yellow).}
    \label{fig:asymm-psuedo-threshold}
\end{figure*}

\subsection{\label{asymm:sub:SymmAsymmNoise}Performance on Symmetric and Asymmetric Noise Models}

The graphs in Fig.~\ref{fig:pseudo} imply that the probability of bit flip error increases with increasing asymmetry of the surface code structure. This naturally raises the question of whether, for asymmetric depolarizing noise, it is suitable to increase only $d_z$ with asymmetry, or should $d_x$ also be increased at a regular interval. In the second experiment, we numerically establish that when the degree of asymmetry is low, one should indeed increase $d_x$ along with $d_z$ to avoid performance degradation. But with a higher degree of asymmetry, it is better to keep $d_x$ fixed and increase $d_z$ only.

We first analyzed how the performance of a set of asymmetric surface codes $[d_x,d_z]$ varies with an increase in asymmetry ($\Delta$) in the depolarizing error channel: $[p_x = p/(\Delta+2), p_y = p/(\Delta+2), p_z = p\Delta/(\Delta+2)]$, where, $p$ is the value of physical error rate for a given physical error $\kappa$. As presented in Table \ref{table:symm-asymm-noise}, for symmetric noise model, i.e., at $\Delta = 1$, asymmetric models perform poorly in comparison to their symmetric counterparts. For example, the pseudo-threshold for [3, 5] is much lower both [3, 3] and [5, 5]. This behavior of asymmetric models is a consequence of their reduced ability to correct for bit flip errors compared to symmetric models, as discussed in the previous subsection. 

Furthermore, as the asymmetry increases step-wise ($\Delta \rightarrow \Delta+1$), we see an improvement in the performance of asymmetric models, as inferred from their increased pseudo-threshold values. This happens because asymmetric codes can now correct overall more number of logical errors as bit flip errors decrease in the creation of logical errors with the simultaneous increase in the contribution of phase flip errors. Additionally, even among the asymmetric models themselves, ones with $d_x=3$ dominate the performance for $\Delta \geq 4$. Therefore, we infer from Table~\ref{table:symm-asymm-noise} that for $\Delta < 4$, the asymmetry in the channel is \textit{weak}, and one should increase $d_x$ regularly with $d_z$ to obtain optimal performance. However, for $\Delta \geq 4$, the asymmetry of the channel becomes \textit{strong} enough to allow an increase in $d_z$ alone, keeping $d_x$ constant, to improve the pseudo-threshold. Therefore, as asymmetry in the error channel increases, or in other words, when the channel becomes \textit{sufficiently asymmetric}, one can achieve much better logical error correction with fewer physical qubits simply by increasing the asymmetry in the surface code. From our numerical results, the \textit{crossover point} appears to be $\Delta = 4$ for our noise model.

Therefore, when the degree of asymmetry is greater than the \textit{crossover point}, our proposed method leads to a better decoding performance while requiring fewer number of qubits. A traditional distance $d$ surface code creates a single logical qubit using $d^2$ data qubits. Whereas, our proposed modification requires $d_x \times d_z$ data qubits to create a single logical qubit. The percentage savings, therefore, is ${(d_z - d_x)}/{d_z} \times 100\%$. For example, when the degree of asymmetry ($\Delta$) is 10, the $(3,7)$ code achieves a percentage increase of $\simeq 48\%$ in the pseudo-threshold with a percentage savings of $\simeq 57\%$ in the number of qubits.

Subsequently, we also note that, for some initial values of $\Delta$, there is a slight improvement in the performance of symmetric surface codes [3, 3], [5, 5], and [7, 7]. However, these gains soon get diminished as asymmetry is increased further in the depolarization channel. This initial increment could be due to a slight dip in net logical errors because the combined decrease in both $\sigma_x$ and $\sigma_y$ errors overpowers an increase in $\sigma_z$ errors.

\subsection{\label{asymm:sub:SymmAsymmCode}Comparison of Symmetric and Asymmetric Surface Codes}

In our third set of experiments, building upon the results from the previous two subsections, we compared the performance of a series of surface codes $[d_x, d_z]$, where $d_x = 3$ and $d_z \in \{3, 5, \ldots, 17\}$ by analyzing the variation of their pseudo-threshold values for $\Delta \in [1, 10, 20, \ldots 100]$. The resulting data has been plotted in Fig. \ref{fig:asymm-psuedo-threshold}, and it shows that there is an evident increase in the performance of error correction in the presence of asymmetric error channels as the asymmetry in surface codes is increased.

We compare these performances to that of the three symmetric surface codes $[d_x, d_z]$, where $d_x = d_z \in \{3, 5, 7\}$. We divide the whole range of asymmetry into three regions: (i) symmetric $(\Delta = 1)$, (ii) intermediate asymmetry $(\Delta \in [10, 100])$, and (iii) strong asymmetry $(\Delta \rightarrow \infty)$. In the first region, we see that that the symmetric surface codes perform much better than asymmetric ones for correcting logical errors, majorly due to the decrease in the capability of asymmetric surface code in solving phase errors, as explained in Section \ref{asymm:sub:ParityPhase}. However, the asymmetric surface codes become more efficient at correcting the logical errors as sufficient asymmetry is introduced in the noise model in the second region. Similar to the trend observed in the previous section for low asymmetry ($\Delta \in [2, 9]$), we see a steady increase in the values of pseudo-thresholds as the asymmetry is increased in the surface code $[d_x, d_z]$, i.e., the value of $d_z$ is incremented while keeping $d_x$ constant. In contrast to symmetric surface codes, where the pseudo-threshold values appear to be saturated right from the beginning, for asymmetric surface codes, we first observe a steady decrease in the improvement of pseudo-threshold values. Only then it appears to be saturating for sufficiently bigger asymmetric surface codes, i.e., the number of data qubits being greater than $50$. 

Overall, once there is a \textit{sufficient} increase in the asymmetry in the noise model, the performance of both symmetric and asymmetric surface codes remains almost constant. A remarkable consequence of this observation is that by using asymmetric surface codes, one can get significantly improved pseudo-threshold rates while requiring less than half the number of physical qubits than the symmetric surface codes. For example, for any $\Delta>10$, we were able to almost double the pseudo-threshold rates with our $[3, 5]$ surface code in comparison to both $[5, 5]$ and $[7, 7]$, decreasing physical qubit counts by $59.18\%$ and $70.10\%$ respectively. Furthermore, there's an observable slight dip in the final region where only phase errors exist as $\Delta \rightarrow \infty$. This slight decrease in this region can be attributed to the fact that the complete absence of any $\sigma_x$ and $\sigma_y$ errors allows for slightly more logical Z errors to come up from the possible ${}^{d_xd_z}C_{d_z}$ combinations.

\begin{table}[t]
\centering
\caption{Variation in threshold ($\gamma^{*}$) values with respect to change in the asymmetry of Pauli errors in depolarizing channel.}
\label{table:asymm-noise-threshold}
\begin{tabular}{|c|c|}
\hline
\textbf{Asymmetry} & \textbf{Thresholds} \\
$(\Delta)$         & ($\gamma^{*})$        \\ \hline
1                  & 0.261               \\ \hline
10                 & 0.278               \\ \hline
20                 & 0.363               \\ \hline
30                 & 0.383               \\ \hline
40                 & 0.396               \\ \hline
50                 & 0.407               \\ \hline
60                 & 0.413               \\ \hline
70                 & 0.423               \\ \hline
80                 & 0.437               \\ \hline
90                 & 0.443               \\ \hline
100                & 0.450               \\ \hline
$\infty$           & 0.500               \\ \hline
\end{tabular}
\end{table}

\subsection{Thresholds values for Asymmetric Surface Codes}

As seen in previous subsections, the relationship between $P_L$ and 
$p$ changes with (i) asymmetric distances of the surface code, represented by $d_x$ and $d_z$, and (ii) asymmetry in the Pauli errors occurring in the channel. In this experiment as well, we have kept $d_x=3$ constant, while varying the $d_z\in\{5, 7, \ldots, 17\}$. For all values of asymmetry $\Delta \in \{1, 10, 20, \ldots, 90, 100, \infty\}$, we see that the values of $P_L$ increases (or decreases) with an increase (or decrease) in $p$ as the $d_z$ is increased until a certain value of $p=\gamma^{*}$ is reached, after which the trend reverses itself. This value $\gamma^{*}$ is known as the threshold rate, and we list them down for asymmetric surface code design in Table \ref{table:asymm-noise-threshold}. Following a similar trend as pseudo-threshold rates, the threshold rates follow a monotonous increasing trend as the asymmetry in the Pauli errors is increased.

\vspace{12pt}
\section{\label{asymm:Conclusions}Conclusions}

This paper investigates a novel design for asymmetric surface codes for the quantum asymmetric Pauli channels. We have proposed the design of $[d_x,d_z]$ surface code, where $d_z > d_x$ in general, for better error correction when the underlying channel is more biased towards Pauli $Z$ errors. In the first look, our rectangular structure ($d_x \cdot d_z)$ seems to use more qubits than the traditional square $(d_x \cdot d_x$) surface code. However, we show that as the asymmetry of the channel increases, our proposed $[d_x,d_z]$ code, where $d_x < d_z$, provides higher pseudo-threshold than traditional surface code, where $d_x = d_z$, using much smaller number of qubits. When the degree of the asymmetry is low, it seems advantageous to somewhat increase $d_x$ regularly with increase in $d_z$ to retain the optimum performance. But with higher degree of asymmetry, we obtain better performance for a constant $d_x$ which is kept fixed at the minimum value of $3$, thus providing percentage savings of ${(d_z - d_x)}/{d_z} \times 100\%$ in qubits. Moreover, by varying $d_x$ instead of $d_z$, these results immediately apply to the inverse case of the noise model studied here, i.e., where bit flips (rather than phase flips) are prevalent. Therefore, we conclude that the proposed asymmetric surface code is more advantageous than symmetric surface codes in the presence of asymmetry in the channel's noise.


\textit{Note Added.} During the revision of this paper, we were made aware of some recent articles \cite{2022arXiv220304948H, 2021arXiv211206036S, 2022arXiv220107802D} which also describe codes tailored to biased noise. These work are based on $XY$ surface codes, $XYZ^2$ stabilizer codes, and Clifford-deformed surface codes, respectively, unlike the $XXZZ$ surface codes we have used in our work.

\section*{Data Availability}

The data that support the findings of this study are available from the corresponding author upon reasonable request.

\section*{\label{asymm:Acknowledgment}Acknowledgement}

This project was started as part of the QIntern 2021 event. Authors acknowledge and thank the QWorld Association for hosting it.


\bibliographystyle{unsrt}
\bibliography{asymmsc}

\end{document}